\documentclass[showpacs,showkeys,preprintnumbers,twocolumn,amsmath,amssymb]{revtex4}

\usepackage{graphicx}% Include figure files
\usepackage{dcolumn}% Align table columns on decimal point
\usepackage{bm}% bold math

\begin{document}
\title{Nuclear Structure Aspects of the Neutrinoless $\beta \beta$-Decays}

\author{E. Caurier $^{*}$, F. Nowacki $^{*}$
and A.~Poves $^{+}$}

\affiliation{({*}) IPHC, IN2P3-CNRS/Universit\'{e} Louis Pasteur
BP 28, F-67037 Strasbourg Cedex 2, France\\
 (+) Departamento de F\'{\i}sica Te\'{o}rica, C-XI. Universidad Aut\'{o}noma
de Madrid, E-28049, Madrid, Spain}

\begin{abstract}
In this article, we analyze some nuclear structure aspects of the 0$\nu$ double beta
decay nuclear matrix elements (NME). We give results for the decays of $^{48}$Ca, 
$^{76}$Ge, $^{82}$Se, $^{124}$Sn, $^{128}$Te, $^{130}$Te, and $^{136}$Xe,
using improved effective interactions and valence spaces. We examine the dependence of
the NME's on the effective interaction and the valence space, and analyze  the
effects of the short range correlations and  the finite size of the nucleon. Finally we study 
the influence of the deformation on the values of the NME's. 
\end{abstract}


\maketitle

\section{Introduction}
\label{sec:intro}

The double beta decay is the rarest nuclear weak process.  It takes place
between two even-even isobars, when the decay to the intermediate
nucleus is energetically forbidden or hindered by the large spin
difference between the parent ground state and the available states in
the intermediate nuclei.  It comes in three forms: The two-neutrino
decay $\beta\beta_{2\nu}$
is just a second order process
mediated by the weak interaction. It conserves the
lepton number and has been already observed in a few nuclei.
The second mode, the neutrinoless decay $\beta\beta_{0\nu}$,
needs an extension of the standard model of the electroweak
interactions as it violates lepton number.  A third mode,
$\beta\beta_{0\nu,\chi}$ is also possible
in some extensions of the standard model and proceeds
via emission of a light neutral boson, a Majoron $\chi$.  The last two
modes, not yet experimentally observed, require massive neutrinos --an
issue already settled by the recent measures by
Super-Kamiokande~\cite{Fukuda.Hayakawa.ea:1998b},
SNO~\cite{Ahmad.Allen.ea:2002a}
 and KamLAND~\cite{Eguchi.Enomoto.ea:2003}.
Interestingly, the double beta decay without emission of neutrinos
would be the only way to sign the Majorana character of the neutrino
and to distinguish between the different scenarios for the neutrino
mass differences.  In what
follows we shall concentrate mostly in the
$\beta\beta_{0\nu}$ mode. The uprising of interest in the
observation of the $\beta\beta_{0\nu}$ decay was somewhat obscured by the
analysis of the status of the calculations of the nuclear matrix elements,
that enter in the lifetime of the decay together with the effective neutrino mass,
made in ref.~\cite{bahcall}. The authors concluded on a very pessimistic note,
saying that the spread of the available values for the nuclear matrix elements
was such that there was no hope to translate the experimental signal into
useful input for the theories beyond the standard model. A critical assessment of the
results of the many calculations available in the framework of the
quasi particle random phase approximation (QRPA) has been made recently \cite{rodin},
with a much more optimistic conclusion. In this article we want to continue
improving upon the reliability of the calculated nuclear matrix elements (NME)
in the framework of large scale applications of the  Interacting Shell Model (ISM).
In addition, we put forward a process of benchmarking the different approaches, and study
the stability of the NME's under reasonable modifications of the nuclear structure
inputs of the calculations.

\section{Double beta decays}

As we have already mentioned, some nuclei, otherwise nearly stable, decay emitting two electrons and two neutrinos
 (2$\nu \; \beta \beta$)
by a second order process mediated by the weak interaction, that has been experimentally measured
in several favorable cases. The decay probability
contains a phase space factor and the square of a nuclear matrix element

\begin{equation}
 [T^{2\nu}_{1/2}]^{-1} = G_{2\nu} |M^{2\nu}_{GT}|^2 
\end{equation}

If the neutrinos are massive Majorana particles, the double beta decay 
can take place without emission of neutrinos  (0$\nu \; \beta \beta$). In this case the transition
is mediated by terms that go beyond the standard model. The decay probability contains
a phase space factor \cite{suci}, the effective electron neutrino mass (a linear combination of the
mass eigenvalues whose coefficients are elements of the mixing matrix) and the nuclear matrix element
\cite{Takasugi:1981,Doi.Kotani.Takasugi:85}.

\begin{eqnarray} 
  [T_{1/2}^{(0\nu)}(0^{+}->0^{+}]^{-1}= \nonumber \\
  G_{0\nu}\left(M_{GT}^{(0\nu)} - \left(\frac{g_V}{g_A}\right)^{2}
   M_{F}^{(0\nu)}\right)^2 \left(\frac{\langle m_{\nu}
   \rangle}{m_{e}}\right)^{2}
\end{eqnarray}

The  2$\nu$ matrix element can be written as:

\begin{equation}
 M_{GT}^{(2\nu)}= \sum_i \frac{\langle GD(J) | \vec{\sigma} t^- | 1_i^+ \rangle 
                      \langle 1_i^+ | \vec{\sigma} t^- | F \rangle }{\Delta E_i}   
\end{equation}

\noindent

The nuclear structure information entering in this matrix element consists on the
wave function of the ground state of the father nucleus (F) and the wave function
of the state of the grand daughter, GD(J), nucleus (J=0, J=1, or J=2)
 to which the decay proceeds. In addition,
the wave functions and excitation energies of all the  1$^+$ states 
in the odd-odd daughter nucleus are in principle necessary.
The spin-isospin operators in the nuclear medium are quenched by a  factor
q$\approx$1/g$_A$, and this quenching factor is also required to reproduce the experimental data
of the  2$\nu$ double beta decays.

To compute exactly the 0$\nu$ matrix element, we would also need the 
 wave function of the ground state of the father nucleus  and of the ground state grand daughter,
 but in addition we would need all the  wave functions and excitation energies of all the 
J$^{\pi}$ states in the odd-odd daughter nucleus.
Most conveniently, the matrix elements can be approximately obtained in the closure approximation, 
                that is good to better than 90\% due
                to the high momentum of the virtual neutrino in the nucleus ($\approx$100~MeV)
Hence, the matrix element can  be written as:

\begin{equation}
M_{GT} ^{(0\nu)} =  \langle GD | h(|\vec{r_1}-\vec{r_2}|) 
(\vec{\sigma}_1\cdot \vec{\sigma}_2) (t^-_1 t^-_2) | F \rangle 
\end{equation}

\noindent
where $ h(|\vec{r_1}-\vec{r_2}|)$ is the neutrino potential ($\approx$1/r).
In this approximation no knowledge of the intermediate nucleus is directly implied.

The transition operators are usually obtained from the Hamiltonian of Doi et al. \cite{Doi.Kotani.Takasugi:85}
                However, according to recent claims \cite{simko},  additional terms originating
                in the coupling to the virtual pions, should give non-negligible contributions.
   The finite size of the nucleon and the short range correlations need to be taken into account
   in the calculation of the two body  matrix elements of the 0$\nu$ two-body transition operators as well.

Perhaps the most relevant issue for our mastering on the neutrinoless double beta decays
is to get a better insight  in the physical content of the two body transition operator.
Only with this knowledge could one decide whether the nuclear wave functions that
enter into the calculation of the NME's contain the degrees of freedom that correspond to
the correlations to which the operator couples dominantly.

The two body decay operators can be written generically as:

\begin{equation}
  \hat{M}^{(0\nu)} = \sum_J  \sum_{i,j,k,l} M_{i,j,k,l}^J \; ((a^{\dagger}_i a^{\dagger}_j)^J \;
                              (a_k a_l)^J )^0     
\end{equation}

\noindent
In words, what the operator does is  to annihilate two neutrons in the parent
                nucleus  and to create two protons The NME is the overlap of the resulting
                object with the grand daughter ground state.

\begin{figure}
 \begin{center}
    \includegraphics[width=1.0\columnwidth]{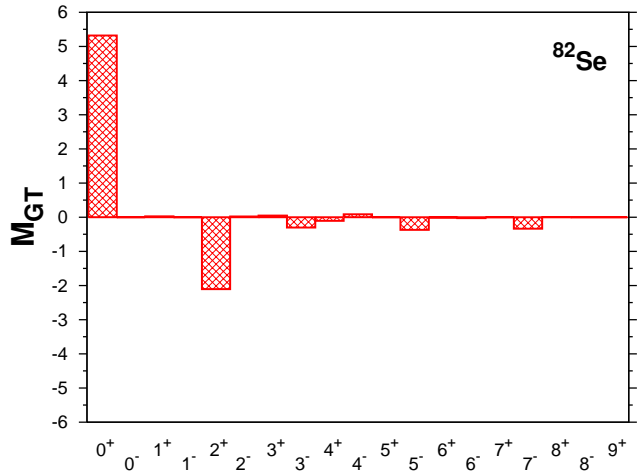}
  \end{center}
\caption{The matrix element of the $^{82}$Se $\rightarrow$ $^{82}$Kr decay as a function of the
     J of the decaying pair \label{fig:se82jj}}
\end{figure}

The contributions to the 0$\nu$ matrix element as a function of the {\bf J} of the
of the decaying pair have a very telling structure as can be seen in Figure \ref{fig:se82jj}.
 The dominant contribution corresponds to the J=0$^+$ pairs, while all the other pairs
have much smaller contributions, but all of them of sign opposite to the leading term.
If the initial and final wave functions had seniority zero, only the leading term would 
contribute and the matrix element would be maximal. This is a first indication, relating
the pairing content of the nuclear wave functions and the neutrinoless double beta 
decay operator.

\section{Interacting Shell Model  (ISM) and  QRPA calculations}

  In the quest for better wave functions to describe the double beta
  decay processes, two main avenues have been explored. The
  interacting shell model in larger and larger valence spaces, with
  ever improving effective interactions, and the quasiparticle
  RPA. Ideally, both methods should be able to produce good
  spectroscopy for parent, daughter and grand-daughter, even better if
  it extends to a full mass region, correct total Gamow Teller
  strengths and strength functions, 2$\nu$ matrix elements, etc. In
  brief, the goal is to have a description as close to perfect as
  possible of the dynamics of the nuclei involved in the
  transition. Until rather recently, ISM calculations that could
  encompass most of the relevant degrees of freedom of the nuclei of
  interest, were only available for the lighter emitters, and, as a
  consequence, most systematic studies were performed in the QRPA framework.
  As we will discuss in the next section, all the
  potential double beta emitters are now within the reach of the ISM
  except $^{150}$Nd, that, being deformed, is also out of the reach of
  the spherical QRPA calculations.

   In the ISM approach, the valence spaces contain a number of orbits
   that is "small" compared to the QRPA, however, all the possible
   ways of distributing the valence particles among the valence orbits
   are taken into account.  In the QRPA calculations, the number of
   active orbits is larger, but only 1p-1h and 2p-2h excitations from
   the normal filling are considered (and not all of them).

	  The effective interactions used in the ISM calculations are
	  usually G-matrices whose monopole behavior is fitted to the
	  spectroscopic properties of a large region of nuclei, in
	  general those comprised between to magic closures for the
	  neutrons and for the protons. In some cases the interactions
	  are plainly fitted to a set of experimental masses and
	  excitation energies. In the QRPA description, the starting
	  point is provided by realistic or schematic interactions,
	  however, both the particle-particle and the particle-hole
	  channels of the interaction are affected of strength
	  parameters dubbed $g_{ph}$ and $g_{pp}$ that are fitted to
	  selected experimental data. In particular, it has been shown
	  that the double beta decay matrix elements depend critically
	  of $g_{pp}$. There are different opinions among the QRPA
	  practitioners about the best choice of the value of
	  $g_{pp}$. The two preferred options being to fit the
	  experimental 2$\nu$ matrix elements, or to fit the Gamow Teller
	  strength functions. In a sense, none is really safe, because
	  the pure $\vec{\sigma} t^{\pm}$ channel plays a rather minor role in the
	  0$\nu$ decay.

  The dominant correlations in the nuclei are due to the pairing and
  quadrupole-quadrupole terms of the nucleon nucleon interaction in
  the medium. In the ISM, the pairing correlations are treated exactly
  within the valence space. Proton and neutron numbers are exactly
  conserved. Proton-proton, neutron-neutron, and proton-neutron
  (iso-vector and iso-scalar) pairing channels are included on equal
  footing. On the contrary, in the spherical QRPA only proton-proton
  and neutron-neutron pairing terms are considered. They are treated
  in the BCS approximation. Proton and neutron numbers are not exactly
  conserved. From the spherical shell model point of view, this
  approach is a seniority zero approximation at the BCS level. 
  When the RPA correlations are taken
  into account, higher seniority components appear in the ground states of the
  father and grand daughter nuclei, however, for the RPA to be valid,
  their amplitudes must decrease rapidly with increasing seniority. 

The multipole correlations and the eventual intrinsic deformation can
be properly described in the laboratory frame in the ISM
description. Angular momentum conservation is exactly preserved.  In
the QRPA the multipole correlations of the ground state are treated at
the RPA level. The posibility of having permanent intrinsic
deformation is not contemplated yet.

\section{The ISM valence spaces relevant for $\beta\beta$ decays}

Classical 0$\hbar \omega$ ISM valence spaces are the $p$, $sd$ and
$pf$ shells. These correspond to the harmonic oscillator major shells
of principal quantum number {\bf p} equal to 1, 2, and 3.  We use
the following notation; in shell {\bf p}, the orbit j=p+1/2 is denoted
by the spectroscopic label of l=p and the remaining ones by r$_p$.
 The space r$_3$g  is
adequate for $^{76}$Ge, $^{82}$Se and their descendants.
With the r$_4$h space for the neutrons and
the g-r$_4$ for the protons, we aim to understand the decays of
$^{96}$Zr, $^{100}$Mo,$^{110}$Pd, and  $^{116}$Cd. Finally the 
r$_4$-h space is a natural choice for the description of the
$^{124}$Sn, $^{128-130}$Te, and $^{136}$Xe decays. 

All these spaces are accessible to large scale ISM descriptions.
The Strasbourg-Madrid \cite{rmp} Shell-Model
codes can deal with problems involving bases of O(10$^{10}$) Slater
determinants, using relatively modest computational resources.

\section{Update of the ISM 0$\nu$ results}
In the valence spaces r$_3$-$g_{9/2}$ ($^{76}$Ge, $^{82}$Se) and
                 r$_4$-$h_{11/2}$ ($^{124}$Sn, $^{128-130}$Te,
                 $^{136}$Xe) we have obtained high quality effective
                 interactions by carrying out multi-parametrical fits
                 \cite{arnaud} whose starting point is given by
                 realistic G-matrices \cite{morten}. With these interactions
                 the beta decay properties of a large set of nuclei are well
                 reproduced. The
                 2$\nu$ double beta decay half-lives are found in reasonably 
                 good agreement with the experimental results as well.
                 In the valence
                 space proposed for $^{96}$Zr, $^{100}$Mo, $^{110}$Pd
                 and $^{116}$Cd, our  results are still preliminary and
                 subject to further improvement both on the interaction side
                 and on the removal of the yet necessary truncations. 

                 Our results are
                 obtained in the closure approximation, with the short
                 range and finite size corrections modeled as
                 described in \cite{0nu1}; using r$_0$=1.2~fm to make
                 the matrix element dimensionless; with g$_A$=1.25;
                 and without higher order contributions to the nuclear
                 current.  A preliminary estimation of the higher
                 order contributions gives a reduction of the ISM
                 NME's in the range of 10\% . Our present best values
                 are collected in table \ref{tab:update}.
$\chi_F$ is defined as:

\begin{equation}
\chi_F =\left(\frac{g_V}{g_A}\right)^{2}   M_{F}^{(0\nu)}
\end{equation}

\begin{table}
\begin{center}
\caption{Update of the ISM  0$\nu$ results\label{tab:update}}
    \begin{tabular*}{\linewidth}{@{\extracolsep{\fill}}llll}
 \hline\noalign{\smallskip}
 \multicolumn{2}{c}{$\langle$m$_\nu$$\rangle$  for T$_{\frac{1}{2}}$ = 10$^{25}$ y.} & { M$^{(0\nu)}_{GT}$} & {1-$\chi_F$} \\
 \noalign{\smallskip}\hline\noalign{\smallskip}
 $^{48}$Ca & 0.85 & 0.67  & 1.14\\
 $^{76}$Ge & 0.90 &  2.35 & 1.10\\
 $^{82}$Se & 0.42 &  2.26 & 1.10 \\
 $^{124}$Sn & 0.45 &  2.11 & 1.13\\
 $^{128}$Te & 1.92 &  2.36 & 1.13\\
 $^{130}$Te & 0.35 &  2.13 & 1.13\\
 $^{136}$Xe & 0.41 &  1.77 & 1.13\\
\noalign{\smallskip}\hline
\end{tabular*}
\end{center}
\end{table}

  Except in the case of doubly magic  $^{48}$Ca,  whose NME is severely quenched, all the other values cluster around a value
   M$^{(0\nu)}$=2.5. The limits of the effective neutrino mass for a half life limit of 10$^{25}$ y, that incorporate the phase space 
    factors, show
   a mild preference for some of the potential emitters.

In table \ref{tab:qrpa} we compare the ISM results with the most
recent QRPA calculations including the higher order corrections
discussed before. The range of values of the NME's shown in the table
is that given by the authors, and derives from the different choices
of g$_{pp}$ and g$_{A}$ used in the calculations, as well as from the
use or not of a renormalized version of the QRPA.  In addition, in
order to make the comparison more transparent, we have selected the
results that treat the short range correlations by means of a Jastrow
factor.  Overall, the two sets of QRPA calculations are now
compatible.  The ISM predictions of the NME's are systematically
smaller than the QRPA central values, except in the case of
$^{136}$Xe.  This nucleus is semi-magic, and the present experimental
limit on its 2$\nu$ decay half life is surprisingly large, perhaps
indicating that some subtle cancelation mechanism is at work.  For the
others, a plausible explanation of the discrepancy, relating it to
the implicit seniority truncations present in the spherical QRPA calculations,
 will be presented elsewhere.

\begin{table}
\begin{center}
\caption{Comparison of the present ISM results with the range of values proposed in 
different QRPA calculations;
(a)  ref.~\cite{rodin,rodin-err}, (b) ref~\cite{kosu}. The values of 
column (a) have been increased by 10\% with respect to the numbers given in 
ref.~\cite{rodin-err}, for a proper comparison with the others, because 
of their  use of r$_0$=1.1~fm. In the same vein, the ISM values should also be
reduced by  $\sim$10\% to account approximately for the absence of higher order 
corrections to the nuclear current.
 \label{tab:qrpa}}
   \begin{tabular*}{\linewidth}{@{\extracolsep{\fill}}lllll}
   \hline\noalign{\smallskip}
 M$_{0\nu}$ &  ISM &  QRPA(a) & QRPA(b)  \\ 
  \noalign{\smallskip}\hline\noalign{\smallskip}
 $^{76}$Ge  & 2.58 & 3.81 - 4.96 & 4.06 - 5.26  \\
 $^{82}$Se  & 2.49 & 3.20 - 4.42 & 2.72 - 3.60  \\
 $^{128}$Te & 2.67 & 2.79 - 4.00 & 3.37 - 4.42  \\
 $^{130}$Te & 2.41 & 2.57 - 3.59 & 3.00 - 4.22  \\
 $^{136}$Xe & 2.00 & 1.39 - 2.32 & 1.92 - 2.92  \\
\noalign{\smallskip}\hline
\end{tabular*}
\end{center}
\end{table}

\section{Exploring the dependences of the ISM results}

\subsection{Effective interaction}

                  The ISM results depend only weakly on the effective
                 interactions provided they are compatible with the
                 spectroscopy of the region. For instance, in the $pf$
                 shell we have three interactions that work properly,
                 KB3~\cite{kb3}, FPD6~\cite{fpd6} and
                 GXPF1~\cite{gxpf1}. Their predictions for the 2$\nu$
                 and the neutrinoless modes are quite close to each
                 other (see table \ref{tab:eff}).  Similarly, in the
                 r$_3$g and r$_4$h spaces, the variations among the
                 predictions of spectroscopically tested interactions
                 are small (10-20\%).

\begin{table}
 \begin{center}
  \caption{Comparison of the  ISM results for the $^{48}$Ca decays using the effective interactions
 KB3,  FPD6, and GXPF1 \label{tab:eff}}
   \begin{tabular*}{\linewidth}{@{\extracolsep{\fill}}llll}   
\hline\noalign{\smallskip}
   &  KB3 & FPD6 & GXPF1 \\
 \noalign{\smallskip}\hline\noalign{\smallskip}
 M$_{GT}$(2$\nu$)  & 0.08 & 0.10 & 0.11\\
 M$_{GT}$(0$\nu$)  & 0.67 & 0.73 & 0.62 \\
\noalign{\smallskip}\hline
\end{tabular*}
 \end{center}
  \end{table}

\subsection{Finite size and short range corrections}

 There has been a certain debate among QRPA practitioners about the amount
of the reduction of the NME's due to the short range correlations
\cite{ucom}. The debate has become milder after the publication of
ref.~\cite{rodin-err}. In the ISM description, once the finite size
of the nucleon is taken into account by means of the standard dipole
form factor \cite{fs}, the effect of the short range correlations,
that we model by a Jastrow ansatz following ref. \cite{src}, is about 
one half of what it would be without it.
In fact, the ISM corrections
for finite size and  short range effects with the Jastrow prescription
are quite close {\bf in percentage} to the QRPA values. 
Short range and finite size corrections proceed
mainly through the reduction of the {\bf J$^{\pi}$=0$^+$} pair contribution.
Preliminary ISM calculations using a softer prescription for the short
range correlations do not change this picture at all. In fact, the relative
values of the ISM NME's with respect to the QRPA ones seem to be independent
of the choice of the (common) prescription used for the  the short range 
correlations.

\begin{table}
\begin{center}
  \caption{Influence of the short range correlations (SR) and of the 
   nucleon finite size (FS) on the nuclear matrix elements of the neutrinoless double beta decays \label{tab:src}}
   \begin{tabular*}{\linewidth}{@{\extracolsep{\fill}}lllll}   
\hline\noalign{\smallskip}
 M$_{GT}$(0$\nu$) & bare & +FS & +SR & +FS+SR   \\
\noalign{\smallskip}\hline\noalign{\smallskip}
  $^{48}$Ca $\rightarrow$$^{48}$Ti    & 1.21 & 0.93 & 0.74 & 0.67   \\
  $^{76}$Ge $\rightarrow$$^{76}$Se    & 3.55 & 2.88 & 2.57 & 2.35   \\
  $^{82}$Se $\rightarrow$$^{82}$Kr    & 3.33 & 2.72 & 2.43 & 2.26  \\
  $^{130}$Te $\rightarrow$$^{130}$Xe  & 3.34 & 2.70 & 2.32 & 2.13   \\
  $^{136}$Xe $\rightarrow$$^{136}$Ba  & 2.76 & 2.24 & 1.93 & 1.77   \\
  \noalign{\smallskip}\hline
\end{tabular*}
\end{center}
\end{table}

\subsection{The influence of deformation}

\begin{figure}
 \begin{center}
   \includegraphics[width=1.0\columnwidth]{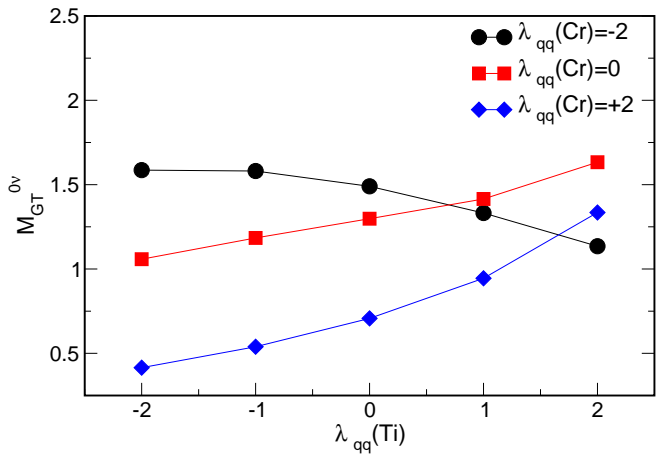}
  \end{center}
 \caption{Influence of deformation in the (fictitious) decay of  $^{48}$Ti.
\label{fig:0nudef}}
   \end{figure}

            Changing adequately the effective interaction used in the
            calculations, we can increase or decrease the deformation
            of parent, grand-daughter, or both, at will. In this
            manner, we can gauge the effect of these variations on the
            decays. We have artificially changed the deformation of
            $^{48}$Ti and $^{48}$Cr adding an extra $\lambda Q\cdot Q$
            term to the effective interaction.  The results are
            presented in Fig.~\ref{fig:0nudef}.  Positive values of
            $\lambda$ increase the deformation while negative values
            reduce it. For zero values of both $\lambda$'s, $^{48}$Cr is already
            well deformed, while $^{48}$Ti is transitional. 
The circle to the upper left corresponds to the spherical-spherical situation,
the square to the upper right to equally deformed Titanium and Chromium, and
the diamond to the bottom left to a spherical Titanium and a very deformed Chromium. Our
            conclusion is that a large mismatch of deformation can reduce
            the $\beta \beta$ matrix elements by factors as large as
            2-3. This exercise indicates that the effect of
            deformation is very important and cannot be overlooked.
            We have reached similar conclusions for heavier emitters
            and a systematic exploration in realistic cases is under
            way.

\section{ISM calculations in QRPA-like valence spaces}

The ISM valence space for the $^{76}$Ge and $^{82}$Se decays has traditionally encompassed the
orbits
1p$_{\frac{3}{2}}$,  0f$_{\frac{5}{2}}$,  1p$_{\frac{1}{2}}$, and  0g$_{\frac{9}{2}}$.
In the QRPA, two major oscillator shells are taken into account;
 0f$_{\frac{7}{2}}$, 1p$_{\frac{3}{2}}$,  0f$_{\frac{5}{2}}$,  
                                                      1p$_{\frac{1}{2}}$,  0g$_{\frac{9}{2}}$,
                              1d$_{\frac{5}{2}}$,  0g$_{\frac{7}{2}}$,
                               2s$_{\frac{1}{2}}$, and 1d$_{\frac{3}{2}}$.

   As a first step toward a more 
                  complete benchmarking, we have evaluated the influence of the 2p-2h jumps 
                  from the 1f$_{7/2}$ orbit -- $^{56}$Ni core excitations-- in our results
                  for the  $^{82}$Se decay. Similar calculations for the  $^{76}$Ge decay are under way.  
   The calculation in the full r$_3$g space plus 2p-2h proton excitations
   from the  0f$_{\frac{7}{2}}$ orbit gives a 20\% increase of M$^{0\nu}$, but probably we overestimate the 
   amount of core excitations. Our   0f$_{\frac{7}{2}}$
   proton occupancies,
   7.71 and 7.69 in  $^{82}$Se and  $^{82}$Kr are smaller than the  BCS 
   occupancies of Rodin et al. \cite{rodin},  7.84 and 7.84.
   Therefore the above 20\% must be taken as a very conservative 
   upper bound. The 2$\nu$ matrix element remains nearly constant, even if the
    total Gamow-Teller strengths, (GT+) and (GT-), increase from  0.15 to 0.34 and from 20.5 to 26.9.
   The larger  GT+ strength is somehow compensated by a larger cancellation
   among the contributions of the different intermediate states and by
   a shift of the centro\"{\i}d of the strength toward higher energy.

   We have also computed the $^{136}$Xe decay in the r$_4$h space including
   2p-2h excitations from the  0g$_{\frac{9}{2}}$proton orbit and
   the matrix element increases less than  10\%.
    In another set of calculations, we have included 2p2h neutron excitations 
             toward the  0h$_{\frac{9}{2}}$  and 1f$_{\frac{7}{2}}$. The occupancies that we obtain are relatively
             large (0.25 neutrons in each orbit) and the effect is to increase the matrix element
             by 15\%. It is interesting to note that the increase with the two orbits simultaneously 
             active is
             equivalent to that obtained including one or another orbit separately. Therefore
             there is no pile-up of the contributions of the small components of the wave function.
    As a preliminary conclusion, the  ISM results seem to be robust against the inclusion 
of small components of the wave function.

\section{Conclusions}

Large scale shell model calculations with high quality effective interactions
                 are available or will be in the immediate future for all but one of the neutrinoless
                 double beta emitters.
The theoretical spread of the values of the nuclear matrix elements entering in the lifetime
                calculations is greatly reduced if the ingredients of each calculation are examined critically
                and only those fulfilling a set of quality criteria are retained.
A concerted effort of benchmarking between ISM and QRPA practitioners 
                 would be of utmost importance to increase the reliability and precision of
                 the nuclear structure input for the double beta decay processes.

\bigskip
{\bf Acknowledgments} This work is supported by the Spanish Ministry of Education and Science with
                a grant FIS2006-1235 and by the IN2P3(France) CICyT(Spain) collaboration agreements.
                Some of the results of this work have been obtained in the framework of the ILIAS 
                integrating activity (Contract R113-CT-2004-506222) as part of the EU FP6 programme
                 in Astroparticle Physics.

\end{document}